\documentclass[12pt]{iopart}
\usepackage{txfonts}
\usepackage{bm}
\usepackage{iopams}
\usepackage{epsfig}
\usepackage{setstack}
\begin{document}
%

\newcommand{\ww}{0.49\linewidth} 
\newcommand{\ttr}{0.32\linewidth} 
\newcommand{\qq}{0.24\linewidth} 
\newcommand{\rhol}{\rho_{\rm loc}}
\newcommand{\vesc}{v_{\rm esc}}
\newcommand{\kpc}{\rm kpc}
\newcommand{\bdisk}{b_{\rm disk}}
\newcommand{\mchi}{m_\chi}
\newcommand{\sigsip}{\sigma_p^{\rm SI}}

\title[Reconstructing WIMP Properties in Direct Detection Experiments]
{Reconstructing WIMP Properties in Direct Detection Experiments Including Galactic Dark Matter Distribution Uncertainties}

\author{Louis E. Strigari}
\address{Kavli Institute for Particle Astrophysics and Cosmology, Stanford University, Stanford, California 94305-4085, USA}

\author{Roberto Trotta} 
\address{Imperial College London, Astrophysics Group, Blackett Laboratory, Prince Consort Road, London SW7 2AZ, UK}


\begin{abstract}
We present a new method for determining Weakly Interacting Massive Particle
(WIMP) properties in future tonne scale direct detection experiments which accounts 
for uncertainties in the Milky Way (MW) smooth dark matter distribution. Using synthetic 
data on the kinematics of MW halo stars matching present samples from the Sloan Digital 
Sky Survey, complemented by local escape velocity constraints, we demonstrate that the 
local dark matter density can be constrained to $\sim 20\%$ accuracy. For low mass WIMPs, 
we find that a factor of two error in the assumed local dark matter density leads to a severely 
biased reconstruction of the WIMP spin-independent cross section that is incorrect at the 
15$\sigma$ level. We show that this bias may be overcome by marginalizing over parameters 
that describe the MW potential, and use this formalism to project the accuracy attainable on WIMP 
properties in future 1 ton Xenon detectors. Our method can be readily applied to different detector 
technologies and extended to more detailed MW halo models. 
\end{abstract} 

\maketitle


\section{Introduction}
The detection of dark matter via elastic scatterings in underground detectors
is a foremost goal of experimental physics in the coming years.  
Upper limits on the spin-independent WIMP-nucleon cross section
~\cite{Ahmed:2008eu,Angle:2008we,Lebedenko:2008gb} are now beginning to carve into
the parameter space of the constrained Minimal Supersymmetric Standard Model (cMSSM)
~\cite{Roszkowski:2007fd,Trotta:2008bp}, which provides a well-motivated theoretical framework for 
the dark matter. Favored MSSSM parameter space 
is expected to be fully probed as future detectors reach the tonne scale. 

A robust interpretation of the experimental limits (or signals) from detectors
requires an understanding of uncertainties associated with the distribution of 
dark matter in the Milky Way (MW), specifically in the vicinity of the solar neighborhood.
Though long regarded a nuisance in the prediction of scattering 
event rates~\cite{Jungman:1995df}, the potential for the uncertainty in the local dark 
matter distribution to bias the reconstruction of the WIMP mass and cross section
has yet to be fully quantified. 

Here we study how well, in a realistic scenario, 
forthcoming direct detection experiments can do in reconstructing the
WIMP mass and cross section, accounting for an uncertain
smooth component of the dark matter distribution, which is independently 
constrained using spectroscopic observations of distant halo stars  
~\cite{Wilkinson:1999hf,Sakamoto:2002zr}
and measurements of the escape speed near the solar circle
~\cite{Smith:2006ym}. We quantify the bias
introduced in the reconstruction of WIMP properties when the
uncertainty in the MW halo model is neglected.  
The results we present are specifically focused
on the spin-independent WIMP-nucleon cross section, but are 
otherwise independent of the specific particle physics framework. 

\section{Milky Way Halo}
Rates in direct detection experiments are proportional to the dark matter density at the 
solar radius. However the potential at the solar radius is likely dominated by baryons, 
so a self-consistent approach must model the combined 
potentials of baryons and dark matter. Disk
stars on circular orbits trace the local potential~\cite{Sumi:2009pu}, 
and the dark matter contribution may be determined via continuation of constraints
at large radii and under the assumption of a smooth dark matter
density profile. This approach is of course subject to systematic uncertainties
in the parameterization of the halo potential, though it 
provides a starting point for understanding the relative contributions
from each of the key MW mass components. As we argue here, even a 
simplified MW model already represents a very considerable improvement in 
bringing under control systematic errors in reconstructed WIMP properties. 

\subsection{Model for Mass Distribution}
We consider a MW halo model that includes  
a central bulge, disk, and dark halo~\cite{Smith:2006ym,Xue:2008se}. 
The bulge is modeled as a spherically-symmetric Hernquist potential,
$\phi_{\rm bulge}(r) = -G \, M_{\rm bulge}/(r+c_0)$, with $c_0 \sim 0.6$ kpc 
and a total mass $M_{\rm bulge} = 1.5 \times 10^{10}$ M$_\odot$.  
The disk surface density is taken of the form
$\Sigma(R) \propto e^{-R/b_{\rm disk}}$,  with $R$ the cylindrical coordinate for
the axially-symmetric disk and its scalelength 
is $b_{\rm disk} \simeq 4$ kpc, which is included as a free parameter in our conservative model (see below). 
The potential from this component is clearly 
non-spherical, though it may be fairly accurately modeled by a spherical 
distribution that has the same mass interior to a Galactocentric radius $r=R$. 
The spherically-symmetric disk potential is taken as 
$ \phi_{\rm disk} = -GM_{\rm disk}(1-e^{-r/b_{\rm disk}})/r$, 
where the total disk mass is $M_{\rm disk} = 5 \times 10^{10}$ M$_\odot$.
 The peak circular velocity for the
 spherical fit to the disk potential at $\sim 2 b_{\rm disk}$ 
 is $\sim 15\%$ less than the circular  velocity of the exponential disk and 
 asymptotes to $\lesssim 5\%$ of the true mass distribution for large radii~\cite{BT1987}. 
For the dark matter halo we take a five parameter model, 
\begin{equation}
 \rho(r) = \frac{\rho_0}{(r/r_0)^a [1+(r/r_0)^b]^{(c-a)/b}}.
 \label{eq:rho} 
 \end{equation}  
 The escape velocity, $v_{\rm esc}$,
 and the circular velocity at the solar radius, $v_0$, are determined from the combined potential 
 of the three components. 
 
As is the case with any parametric model, 
the results we present will likely vary if the model is not an accurate 
description of the true Galaxy. As such, we view our analysis as 
means of estimating uncertainties and bias on key dark matter parameters
for a well-defined, though perhaps simplified, MW model, and acknowledge that the above 
model provides a first step towards consideration of a wider range of Galactic models
in this context. For example, to better model observations~\cite{Merrifield:2003hz} triaxial 
models for the MW halo may be considered which increase the dark matter 
density in the disk. However, on the theoretical side there are uncertainties to
the predicted MW halo shape, e.g. simulations suggest that adding gas cooling
 tends to make halos more spherical~\cite{Kazantzidis:2004vu}. 
 
\subsection{Stellar Kinematics} 
From the above parametric model for the Galaxy, we simulate 
stellar kinematics data which we then employ to determine the accuracy with which the particle properties of the dark matter 
and its astrophysical distribution can be reconstructed. Our synthetic data sample consists of $2000$ stars distributed uniformly
with Galactocentric radii in the range $5-40$ kpc, similar to the distribution 
in recent Sloan Digital Sky Survey measurements~\cite{Xue:2008se}. 
The distance, $d$, from a halo star to the Sun is given by 
$d^2=r^2+r_\odot^2-2\,r\,r_\odot\, \cos \theta$, so that the line-of-sight velocity 
to the star is $\dot d = \dot r (r/d-r_\odot \cos \theta/d)+r\, r_\odot \,\dot \theta \sin \theta/d$. 
Here $\theta$ is the spherical polar angle and 
$r_\odot=8.5$ kpc is the distance from the Sun to the Galactic center, which 
we keep fixed. 
The line-of-sight velocity dispersion is defined as the average of $\dot d^2$
over the solid angle in Galactic coordinates, 
$\sigma_{\rm los}^2 \equiv \langle \dot d^2 \rangle$. 
Performing this average gives~\cite{Dehnen:2006cm},
\begin{equation}
\sigma_{\rm los} = \sqrt{1-A(r)\beta(r)} \sigma_r, 
\label{eq:sigma_los}
\end{equation} 
where
\begin{equation} 
A(r) = \frac{r^2+r_\odot^2}{4r^2} - \frac{(r^2-r_\odot^2)^2}{8r^3r_\odot}
\ln \left | \frac{r+r_\odot}{r-r_\odot} \right |. 
\label{eq:Ar}
\end{equation}
The velocity anisotropy parameter is 
$\beta(r) = 1-\sigma_\theta^2/\sigma_r^2$, which we assume 
to be constant as a function of radius. 

Assuming that the dark matter halo and the tracer population of halo 
stars are in equilibrium, 
the radial jeans equation for the intrinsic, or statistical, dispersion $\sigma_r$ is 
\begin{equation}
\frac{d (\rho_\star \sigma_r^2)}{dr} + \frac{2\rho_\star\sigma_r^2 \beta(r)}{r} 
= -G\rho_\star\frac{d\phi}{dr}, 
\label{eq:jeans}
\end{equation}
where $\rho_\star(r)$ is the spatial density profile for the tracer population. 
The potential on the right-hand side
is the sum of the potential from the disk, bulge, and the dark matter, 
$\phi = \phi_{\rm disk}+ \phi_{\rm bluge} + \phi_{\rm dm}$. 
The density of halo stars is modeled to fall off 
according to  $\rho_\star \propto r^{-3.5}$ beyond $10$ kpc and remain 
constant for  $r \le 10$ kpc~\cite{Bell:2007ts}.  
  
\section{Direct Detection}
To calculate WIMP-nucleon recoil event rates, we must estimate both the local WIMP density and the 
velocity distribution of the dark matter. For the latter, we appeal to recent
results from high resolution numerical simulations, which find only mild deviations from an isotropic and Maxwellian velocity 
distribution in the solar neighborhood~\cite{Hansen:2005yj,Vogelsberger:2008qb,Fairbairn:2008gz}. This implies 
10\% deviations in recoil event rates relative to the 
standard isotropic Maxwellian assumption~\cite{Evans:2005tn}. 
Motivated by these results, we assume 
an isotropic Maxwellian form for the dark matter velocity distribution, characterized by 
$v_{\rm esc}$, $v_0$, and local density
$\rho(8.5 \, {\rm kpc})$~\cite{Lewin:1995rx}, which are all computed from our MW model 
described above and constrained as explained below.

We consider the case of a Xe target, and assume
everywhere a 1 tonne-yr exposure. 
In order to model the energy resolution of the detector, the recoil energy spectrum is smoothed with an energy-dependent 
Gaussian window function of standard deviation $\sigma_E(E) = 1.5\sqrt{E}$. For definiteness, our results below are obtained by considering 5 
equally-spaced energy bins within the recoil range $5-30$ keV, 
and adopting an independent Poisson likelihood in each bin, with negligible background counts. However, our results are insensitive to the details of the binning scheme and to the precise choice of the upper cutoff in the recoil spectrum. The nuclear form factor 
for spin-independent scatterings is taken to be the
standard Helm form~\cite{Lewin:1995rx}. For the sake of simplicity, we 
do not include the yearly variation in the event rate resulting from the projection
of the Earth's orbital velocity on the Galactic plane; we find that including this effect 
has negligible effect on the results we present. 

\section{Statistical methodology}
We consider the following fiducial parameters for the Milky Way: $\Theta = \{\rho_0,r_0,\beta,a,b,c,b_{\rm disk}\}$ = 
\{$7.4 \times 10^6$ M$_\odot$ kpc$^{-3}$, 20 {\rm kpc}, 0, 1, 1, 3, 4 {\rm kpc}\}. This set of 
dark halo parameters is broadly motivated by numerical simulations of Milky Way mass halos
~\cite{Navarro:1996gj}, and as discussed above the mass and scalelength of the disk, $b_{\rm disk}$, 
are of the order observed for the Milky Way~\cite{Widrow:2008yg}. 
For this set of parameters the local dark matter 
density is 0.32 GeV cm$^{-3}$, the escape velocity $v_{\rm esc} = 547$ km s$^{-1}$ and the circular 
velocity $v_0 = 209$ km s$^{-1}$. We find that the results presented below are insensitive to
similar sets of fiducial model parameters, provided that they roughly match both the local 
dark matter density and the mass and circular velocity of the Milky Way. 
The likelihood is centered around the fiducial values, and is the product  of three 
components: the projected direct detection constraints, the projected stellar kinematics constraints, and present-day 
escape velocity measurements. 

We define the likelihood for the projected velocity component along the line-of-sight to a given star 
to be Gaussian with zero mean and a variance given by the 
$\sigma_{{\rm los}}$. The total likelihood is then obtained by  
multiplication of the Gaussian likelihood for each star. 
The dispersion in the likelihood is dominated by the intrinsic 
dispersion at a given radius, i.e.\ we ignore contributions to the
likelihood that involve uncertainty on the measured velocity 
of each star.  Assuming that the model parameters 
only enter through the velocity dispersion, 
it is straightforward to derive the 
Fisher information matrix, which is defined as the second derivative of the log of the 
likelihood function, with respect to the model parameters, 
$\langle \partial^2 {\cal L} /\partial \Theta_a \partial \Theta_b \rangle$.
For our likelihood the Fisher matrix is simply given by 
\begin{equation} 
F_{ab} 
= \sum_{\imath=1}^{N}
\frac{1}{2}
\frac{1}{\sigma_{\rm los,\imath}^4}\frac
{\partial \sigma_{\rm los,\imath}^2}{\partial \Theta_a}
\frac{\partial \sigma_{\rm los,\imath}^2}{\partial \Theta_b}.
\label{eq:fisher}
\end{equation} 
Written in this form we assume that the mean velocity does not
depend on our model parameters; this is likely to be a good 
assumption if we consider that the intrinsic dispersion of the 
halo stars dominates the contribution due to ordered rotational
motion. 

We take the constraint on the escape velocity to be Gaussian with mean 
$\bar{v}_{\rm esc} = 544$ km s$^{-1}$ and standard deviation 
$\sigma = 33$ km s$^{-1}$~\cite{Smith:2006ym}. For simplicity 
we do not consider additional constraints on the MW mass
distribution that come from the inner and outer rotation curves or from the 
motions of stars in the solar neighborhood, as considered in Refs.
~\cite{Dehnen:1996fa,Widrow:2008yg,Catena:2009mf}. 
Our motivation for including only the halo stars and the escape velocity 
constraints in our data set is that these quantities are the most direct
tracers of the MW dark matter halo. Given these inputs, 
the projected constraints on both the local dark matter density and the 
reconstructed of WIMP properties we present below should be viewed as a 
conservative estimate of the errors attainable on these parameters. 

We adopt a Metropolis-Hastings algorithm to perform a Markov chain Monte Carlo scan over the parameters defining the MW halo model described above, augmented by the WIMP mass $m_\chi$ and its spin-independent scattering cross section $\sigma_p^{SI}$. We assume flat priors on all MW halo parameters over the ranges 
$\{ \log [\rho_0/({\rm M}_\odot {\rm kpc}^{-3})] ,r_0/{\rm kpc}, \beta, a,b,c,b_{\rm disk}/{\rm kpc} \} = \{[5:8],[10:60], [-1:1], [0:2],[0:2],[2:4],[2:6]\}$, as well as flat priors over sufficiently wide ranges of $\log m_\chi$ and $\log\sigma_p^{SI}$. We find that our results are insensitive to the details of the prior choice. 

\begin{figure}
\begin{center}
\begin{tabular}{c}
\includegraphics[width=0.65\linewidth]{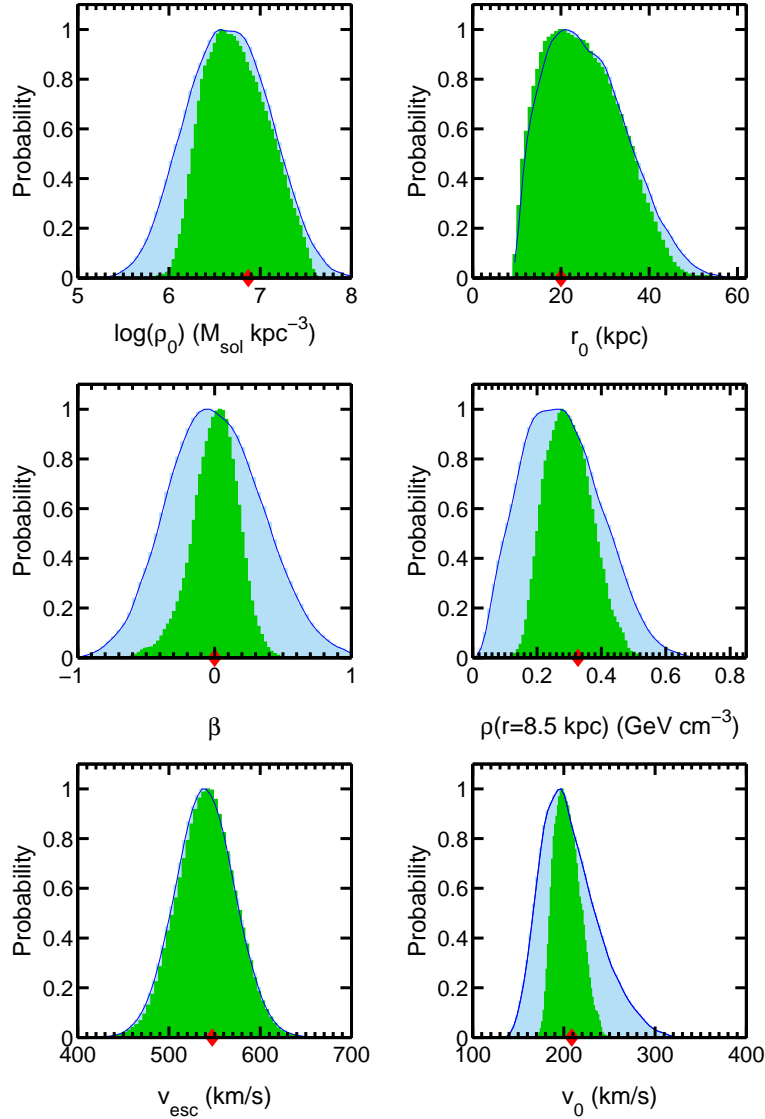}
\end{tabular}
\end{center}
\caption{Reconstruction of 3 halo model parameters ($\log\rho_0, r_0, \beta$) and constraints on WIMP scattering rates-related quantities at the solar circle ($\rho(8.5~{\rm kpc}), v_{\rm esc}, v_0$) using 2000 halo stars and a local determination of the escape velocity. 
Green (tighter) and blue (wider) curves show the results for the 3 parameter ({\em baseline})
and 7 parameter ({\em conservative}) halo model. The red diamond is the true value.
 }
\label{fig:marginal}
\end{figure}

\section{Results}  
We study three representative cases, which differ in the 
number of halo parameters that are marginalized over. Our {\em fixed}
model takes $\Theta$ to be fixed to an assumed value, which might not match the correct one. This corresponds to neglecting all astrophysical uncertainties while not including any of the astrophysical constraints.
Our {\em baseline} model varies the following three parameters: $\{\log \rho_0 ,r_0,\beta\}$, which are constrained using the astrophysical data and marginalized over when considering direct detection constraints, while the remaining four parameters are fixed to their fiducial value. In addition to the three parameters above, our {\em conservative} model also 
marginalizes over the remaining four parameters of the halo model, 
$\{a,b,c,b_{\rm disk}\}$. 

Figure~\ref{fig:marginal} shows the 1D posterior marginal probabilities for 
several parameters for the baseline and the conservative models, after the halo stars and escape velocity constraints have been applied.  The distributions are correctly peaked around the true value and in particular the density of dark matter at 8.5 kpc can be constrained to 22\% (37\%) at $1\sigma$ in the baseline (conservative) model. We find that the local escape velocity constraint is important in tightening the bounds on the WIMP local density. 

\begin{figure*}
\includegraphics[width=0.45\linewidth]{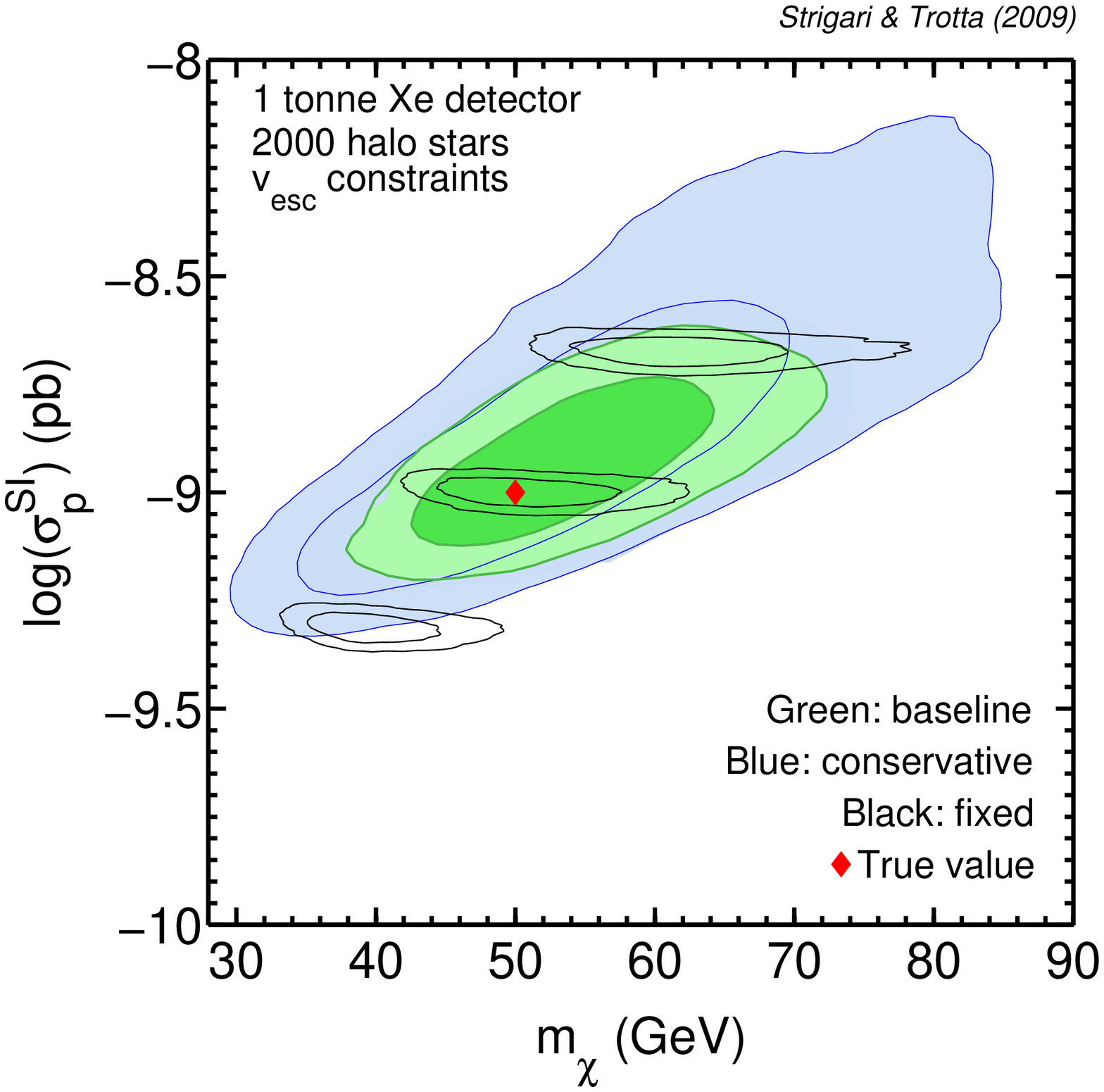} \hspace{2cm} 
\includegraphics[width=0.45\linewidth]{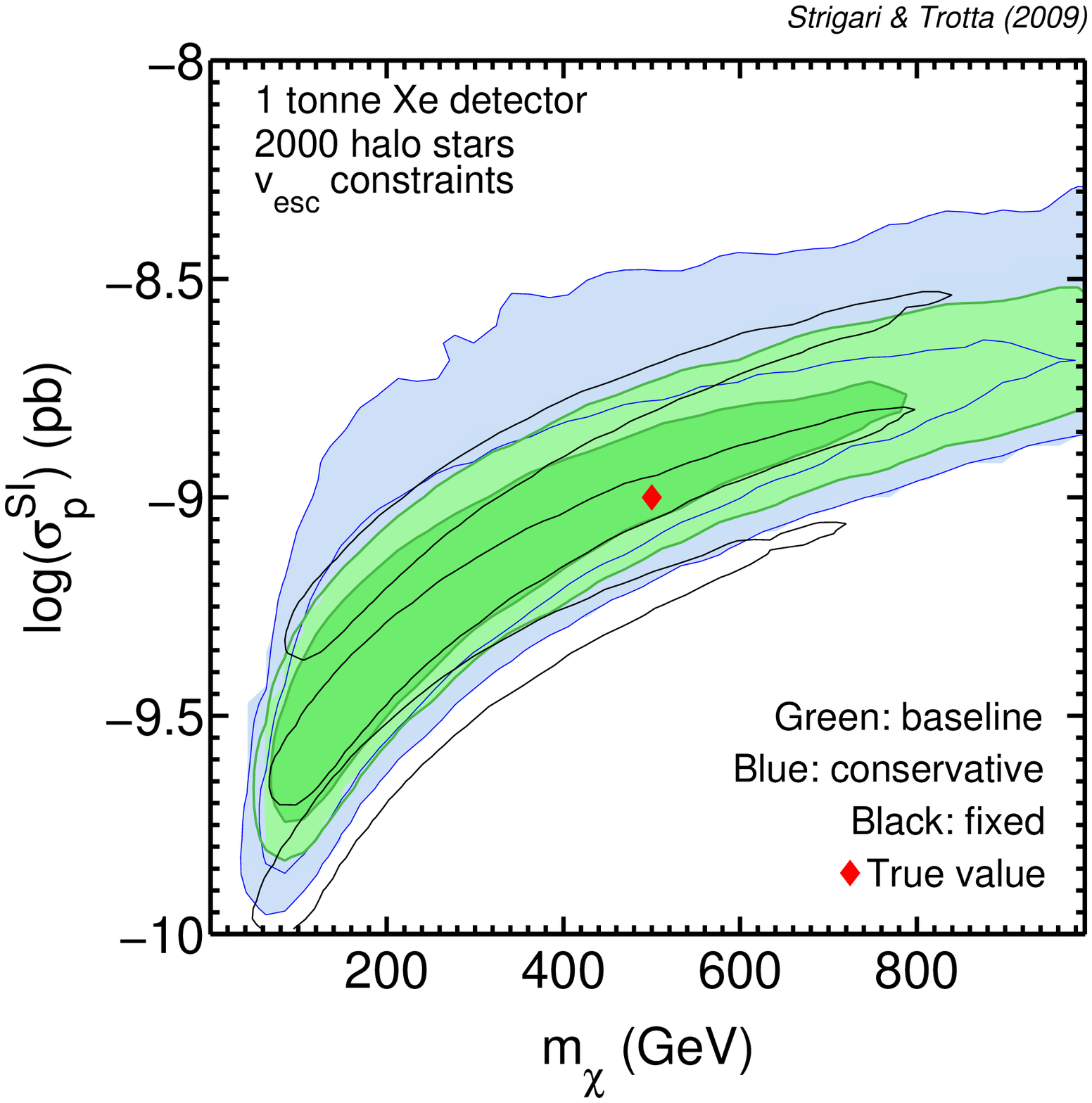}
\caption{Reconstruction of $m_\chi, \sigma_p^{SI}$ under 
various assumptions: dark matter halo parameters fixed to 
assumed values (solid, black), 
marginalizing over baseline halo model (filled/green), 
marginalizing over conservative  halo model (filled/blue). In all cases inner and outer contours
represent 68\% and 95\% c.l. limits. The red diamond gives the true value. The left panel is for a 50 GeV WIMP mass, 
the right panel assumes a 500 GeV WIMP (In the right panel, we only show 68\% c.l. for the case of fixed galactic parameters for clarity).
The lower (upper) solid black contours illustrate the bias in the reconstruction assuming incorrect values for the local 
dark matter density a factor of 2 above (below) the true value.}
\label{fig:bias}
\end{figure*}

Figure~\ref{fig:bias} depicts the reconstruction of $m_\chi$ and $\sigma_p^{SI}$ 
for a 1 tonne-yr Xe exposure for the baseline (filled/green) and conservative (filled/blue) models, assuming a 
fiducial value $\sigma_p^{SI} = 10^{-9}$ pb.
We show the case of a low mass ($m_\chi = 50$
GeV, left panel) and a high mass WIMP ($m_\chi = 500$ GeV, right panel), to highlight the difference
between WIMP mass scales that are well-constrained and poorly constrained~\cite{Bernal:2008zk,Green:2008rd,Shan:2009ym}. 
At $m_\chi = 50$ GeV, the degeneracy between $m_\chi$ and $\sigma_p^{SI}$
is broken by recoil energy spectral information; we find that a minimum of 3 evenly
spaced bins suffices to break the degeneracy.  The fiducial model with $m_\chi = 50$ GeV ($m_\chi = 500$ GeV) 
produces 383 (79) events within our energy window for a tonne-yr exposure for Xe. 

Figure~\ref{fig:bias} also shows the bias in 
the reconstruction resulting from assuming the incorrect local dark matter density. 
For a 50 GeV WIMP, fixing the local dark matter density a factor of two above or 
below the true value biases the reconstruction
of the WIMP cross section by $\sim 15\sigma$. 
The effect is less severe for a 500 GeV WIMP, of order $\sim 5 \sigma$, though this case is intrinsically much less
constrained. This systematic bias would {\em not} be detectable by the usual goodness-of-fit test, 
for in all three cases depicted in Figure~\ref{fig:bias} the reduced chi-square of the best-fit point is  statistically
indistinguishable. Figure~\ref{fig:bias} clearly 
shows the necessity of accounting for uncertain astrophysics in the
determination of $m_\chi, \sigma_p^{SI}$, even in a simplified 
MW halo model. For the well constrained 50 GeV WIMP case, marginalizing over the halo model parameters correctly recovers the true point with no bias, while
the error in the mass increases from 9\% to 14\% (24\%) for the baseline (conservative) model. The impact on the accuracy on $\log\sigma_p^{SI}$ is more severe, increasing the relative 
error by a factor of 5 under the baseline model and by a factor of  20 for the conservative case. However, this 
reduced statistical accuracy is compensated by a greatly increased robustness to systematic bias. 

\section{Conclusion}
We have presented an analysis of the constraints attainable on 
WIMP parameters using future direct detection experiments. Our 
key results are twofold: first we have quantified how well WIMP properties can 
be constrained for an optimistic particle mass in tonne scale detectors, 
and second we have demonstrated that including an explicit model of the MW halo can overcome a potentially severe ($\sim 15\sigma$) bias in the reconstruction of WIMP
parameters when the wrong fiducial values for the halo are assumed. 
In future analyses
our MW halo model can be refined to include triaxial shapes for the 
dark and luminous components.
The contribution from dark matter substructure~\cite{Kamionkowski:2008vw,Vogelsberger:2008qb}
or a rotating component~\cite{Read:2009jy} may be considered. 
Further one may account for the non-Maxwellian velocity distribution 
~\cite{Peter:2009mi}, and a multi-component spectral fit to  
the WIMP and astrophysical background spectra may be incorporated
~\cite{Monroe:2007xp,Strigari:2009bq}. 
An analysis along these lines will be crucial to  
interpret the limits and measurements from forthcoming direct 
detection experiments.  

{\bf Acknowledgments}--
We thank Henrique Araujo, Laura Baudis, Blas Cabrera, Jodi Cooley-Sekula, and Alastair Currie for several important
discussions. Support for LES was provided by NASA through Hubble Fellowship grant 
HF-01225.01 awarded by the Space Telescope Science Institute, which is 
operated by the Association of Universities for Research in Astronomy, Inc., 
for NASA, under contract NAS 5-26555. RT would like to
thank the EU FP6 Marie Curie Research and
Training Network ``UniverseNet'' (MRTN-CT-2006-035863) for partial
support. 

\bibliography{paper}

\end{document}